\documentclass[twocolumn,showpacs,preprintnumbers,amsmath,amssymb]{revtex4}

\usepackage{graphicx}
\usepackage{dcolumn}
\usepackage{bm}
\usepackage{epsfig}

\begin{document}


\title{Implementing Non-Projective Measurements via Linear
Optics:\\
an Approach Based on Optimal Quantum State Discrimination}

\author{Peter van Loock$^1$, Kae Nemoto$^1$, William J.\ Munro$^{1,2}$,
Philippe Raynal$^3$, and Norbert L\"{u}tkenhaus$^3$}

\affiliation{ $^1$National Institute of Informatics, 2-1-2
Hitotsubashi, Chiyoda-ku, Tokyo 101-8430, Japan\\
$^2$Hewlett-Packard Laboratories, Filton Road, Stoke Gifford,
Bristol BS34 8QZ, United Kingdom\\
$^3$Quantum Information Theory Group, Institute of Theoretical
Physics I, Institute of Optics, Information and Photonics
(Max-Planck Forschungsgruppe), Universit\"{a}t
Erlangen-N\"{u}rnberg, Staudtstr.7, D-91058 Erlangen, Germany
}


\begin{abstract}
We discuss the problem of implementing generalized measurements
(POVMs) with linear optics, either based upon a static linear
array or including conditional dynamics. In our approach, a given
POVM shall be identified as a solution to an optimization problem
for a chosen cost function. We formulate a {\it general
principle}: the implementation is only possible if a linear-optics
circuit exists for which the quantum mechanical optimum (minimum)
is still attainable after dephasing the corresponding quantum
states. The general principle enables us, for instance, to derive
a set of necessary conditions for the linear-optics implementation
of the POVM that realizes the quantum mechanically optimal
unambiguous discrimination of two pure {\it nonorthogonal} states.
This extends our previous results on projection measurements and
the exact discrimination of orthogonal states.
\end{abstract}

\pacs{03.67.Hk, 42.25.Hz, 42.50.Dv}

\maketitle

\section{Introduction}

The implementation of {\it positive operator-valued measures}
(POVMs) for photonic quantum state signals is an essential task in
many quantum information protocols. In general, in order to
implement such measurements, a nonlinear interaction of the signal
states, described by a Hamiltonian at least cubic in the optical
mode operators \cite{LloydBraunstein}, is needed. With current
technologies, however, these nonlinear effects are hard to obtain
on the level of single photons. Apart from hybrid schemes based on
weak nonlinearities and strong coherent probe pulses
\cite{Nemoto}, an alternative approach for inducing a nonlinear
element is to exploit the effective nonlinearity associated with a
measurement. In particular, for photonic-qubit states, universal
quantum gates and hence any POVM can be realized deterministically
or asymptotically (near-deterministically), using linear optics,
photon counting, entangled auxiliary photon states, and
conditional dynamics (feedforward)
\cite{KLM01,FransonPRL,Nielsen,BrowneRudolph}. Moreover, cheaper
resources may suffice for the implementation of non-deterministic
gates and POVMs, using feedforward \cite{KLM01,FransonPRA} or a
static array of linear optics
\cite{KLM01,Knill1,Knill2,Timcoincide,ScheelKaeBill,ScheelNorbert,Jens04}.
Here we will focus on the implementation of POVMs using either
static linear optics or feedforward and, in particular, photon
counting. Although there are some specific results on this issue
\cite{CalsamPOVMpaper}, a general and practical solution to the
problem as to whether a given POVM can be implemented by linear
optics is not known. Only for the special class of projective
measurements, a set of simple criteria has been derived
\cite{PvLNorbert}.

In the special case of a {\it projection measurement}, the
``signal states" to be distinguished (i.e., the basis that spans
the space to be projected on) are orthogonal. In this case,
quantum mechanically, an exact discrimination with unit
probability for a conclusive result is possible. However, if the
implementation of the projection measurement is restricted to a
limited class of transformations such as passive linear optics or
Gaussian transformations, unit probability might be unattainable
\cite{PvLNorbert,PvLPhilippeNorbert}. The prime example to which
such a no-go statement applies is the Bell measurement for
polarization-encoded photonic qubit states
\cite{Vaidman99,NL99,PvLNorbert,PvLPhilippeNorbert}. Of course,
such a no-go statement for {\it exact} state discrimination does
not rule out the possibility for near-deterministic or
non-deterministic implementations. For example, the simplest
approximation to the single-photon qubit Bell measurement only
requires a symmetric beam splitter and photon counting. This
scheme achieves a success probability of one half, thus attaining
the upper bound when using linear optics and photon counting, but
neither auxiliary photons nor feedforward \cite{Calsam01}.

A hierarchy of simple criteria for the exact discrimination of
orthogonal states can be derived via a {\it dephasing approach}
\cite{PvLNorbert}. The idea of this approach is to simulate the
actual detection, for instance, in the photon number basis through
a dephasing of the linearly transformed states, turning them into
mixtures diagonal in the Fock basis. Any term in these mixtures
represents a possible detection pattern for a given input state
and a given linear-optics circuit. The requirement for an exact
discrimination of the signal states is then that the overlap of
the dephased density operators vanishes, corresponding to the
non-existence of any coinciding patterns. Expressing the overlap
in terms of the fidelity, this means that the fidelity of the
orthogonal states must remain zero after the linear transformation
and the dephasing operation have been applied to the states.

In order to extend the analysis of projection measurements
\cite{PvLNorbert} to {\it generalized measurements}, the first
obvious approach is to consider von Neumann measurements in a
larger Hilbert space. Suitably chosen, these are then equivalent
to the POVM in the smaller signal space. In fact, any POVM can be
expressed in such a way via the Naimark extension. For signal
states having only {\it one photon}, already the Naimark extension
approach reveals that {\it any} POVM can be implemented with
linear optics. A demonstration of this can be found in
App.~\ref{onephoton}. Recent theoretical work on linear-optical
implementations of one-photon POVMs and Kraus operators can be
found in Refs.~\cite{Ahnert,Ahnert2}. Previously, one-photon POVMs
via linear optics, in particular, for quantum state discrimination
were considered in Refs.~\cite{Bergou00,Sun01,Sun02}. There have
been also several experimental linear-optics realizations of a
non-projective one-photon POVM, namely that for unambiguous state
discrimination \cite{Mohseni04,Huttner96,Clarke01} (a review of
experimental state discrimination can be found in
Ref.~\cite{Cheflesreview}). In general, however, for signal states
with arbitrarily many photons, to decide whether an exact
implementation of a given POVM is, in principle, possible with
linear optics is a non-trivial problem. Nevertheless, approximate
two-photon POVMs have been implemented already via linear optics,
for instance, for realizing a ``nonlocal measurement" on a
two-photon state \cite{Pryde} using a non-deterministic two-photon
controlled-NOT gate \cite{Obrien}.

Here, in order to address the question of the implementability of
a general multi-photon POVM with linear optics,
we refer to a {\it fundamental principle}, independent of the
Naimark extension. In order to apply this principle, first, the
POVM shall be identified as a solution of an optimization problem
for some cost function. In terms of this cost function, the
principle then states that the implementation is only possible if
a linear-optics circuit exists for which the quantum mechanical
optimum (minimum) is still attainable {\it after dephasing} the
corresponding quantum states. Whether linear optics or more
general linear transformations including multi-mode squeezing are
sufficient to implement the corresponding POVM depends on the
ability of these tools to obey the above general rule. Applying
this rule to the fidelity of two nonorthogonal states will enable
us to derive a set of necessary conditions for the implementation
of the quantum mechanically optimal unambiguous state
discimination (USD), extending our analysis of discriminating
orthogonal states \cite{PvLNorbert}. The USD of non-orthogonal
states is a simple example for a non-projective POVM, where some
measurement results are inconclusive, but the remaining results
correctly identify the signal state.

The plan of the paper is as follows. First, in
Sec.~\ref{Secdephase}, we are going to explain how the effect of
the detection behind a linear-optics circuit can be described via
dephasing. This enables us to present the main result of the
paper, a general principle for the implementation of POVMs with
linear optics. In Sec.~\ref{Secprojec}, we briefly review how the
known criteria for linear-optics projection measurements follow
from this general principle as a simple special case. Finally, we
turn to the implementation of non-projective POVMs in
Sec.~\ref{Secnonprojec}, where our main focus will be on the
unambiguous discrimination of two pure non-orthogonal states.

\section{The dephasing approach to POVMs}\label{Secdephase}

Given a general non-projective POVM, via the Naimark extension
approach it is pretty hard to decide whether the POVM can be
implemented with linear optics.
Here we propose an alternative strategy independent of the Naimark
extension, based upon a dephasing approach. The dephasing effect
will be used to mimic the projection of the individual modes onto
the detection basis. Let us first introduce the dephasing
formalism.

The dephasing basis is determined by the detection mechanism of
the implementation. This might be either the discrete photon
number basis (photon counting) or the continuous quadrature
eigenstate basis (homodyne detection). In the following, we will
use the Fock basis as the dephasing basis. This basis can be
easily substituted by other appropriate bases \cite{PvLNorbert}.

If the signal states $\hat\rho$ are linearly transformed into the
states $\hat\rho_H$, and the photon number of the modes will be
detected, the corresponding dephasing effect can be described as
\begin{eqnarray} \label{dephasingmechanism}
\hat\rho_H\rightarrow \hat\rho_H'=\frac{1}{(2\pi)^N} \int d\phi^N
e^{-i\vec a^\dagger D \vec a}\hat\rho_H e^{i\vec a^\dagger D \vec
a}\;.
\end{eqnarray}
Here, we used $d\phi^N\equiv d\phi_1d\phi_2 ... d\phi_N$, the
diagonal $N\times N$ matrix $D$, $(D)_{ij}=\delta_{ij}\phi_i$, and
the vectors $\vec a=(\hat a_1,\hat a_2, ... ,\hat a_N)^T$ and
$\vec a^\dagger=(\hat a_1^\dagger, \hat a_2^\dagger, ... ,\hat
a_N^\dagger)$, representing the annihilation and creation
operators of all the electromagnetic modes involved.

\begin{figure}[tb]
\epsfxsize=1.2in \epsfbox[220 330 400 560]{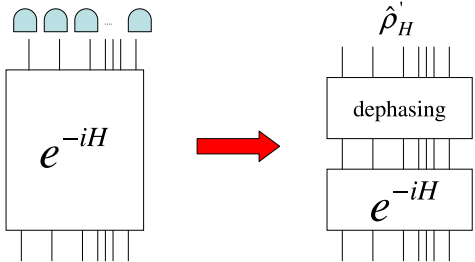}
\caption{\label{fig1} (Color online) An equivalent description of
the detection mechanism after a unitary state transformation via
dephasing. In the case of photon counting, the dephased density
matrix is a mixture of all possible photon number patterns for a
given input state and a given state transformation. Here, we are
mainly concerned about linear-optics transformations.}
\end{figure}

The effect of the dephasing is that it turns the linearly
transformed states into a Fock-diagonal density matrix. This
mixture contains all possible photon number patterns for a given
input state and a given linear-optics circuit. The weights of the
different terms in this mixture are determined by the
probabilities for obtaining the corresponding pattern via photon
detection. Thus, the dephasing formalism is an equivalent
description for the effect of the detection after the
linear-optics transformation (see Fig.~\ref{fig1}). An example for
a pure signal state $\hat\rho = |\chi\rangle\langle\chi|$, and
hence a pure transformed state $\hat\rho_H\equiv
|\chi_H\rangle\langle\chi_H|$, would be $|\chi_H\rangle = \alpha
|110\rangle + \beta |101\rangle + \gamma |002\rangle$. In this
case, the dephased state becomes $\hat\rho_H' = |\alpha|^2
|110\rangle\langle 110| + |\beta|^2 |101\rangle\langle 101| +
|\gamma|^2 |002\rangle\langle002|$, corresponding to the possible
detection patterns $110$, $101$, and $002$.

The advantage of the dephasing formalism is that the effect of the
detection is described on the level of the state transformations
and the final states become as classical as they can get. These
close-to-classical states can then be analyzed with respect to a
given quantum information task. One may also consider only
partially dephased states which are Fock-diagonal only with
respect to the dephased modes. Partial dephasing mimics those
protocols where only a subset of the modes is detected and a
subsequent linear-optics transformation is applied to the
remaining modes conditioned upon the measurement outcomes
(conditional dynamics).

Using the dephasing formalism, we now propose the following
strategy in order to decide whether a given POVM can be
implemented via linear optics. First, the POVM shall be identified
as a unique solution to an optimization problem. For the cost
function to be optimized, we then refer to a general principle:
the implementation is only possible if a linear-optics circuit
exists for which the quantum mechanical optimum (minimum) is still
attainable {\it after dephasing} the corresponding quantum states.
Thus, optimizing the cost function for the dephased states must
yield the same minimum as for the original signal states. The
linear-optics circuit must be chosen such that
\begin{equation}\label{generalprinciple}
C_{\rm linear\; optics,\; dephasing}^{\rm
optimal}\stackrel{!}{=}C_{\rm quantum\; mechanics}^{\rm optimal}
\;,
\end{equation}
where the symbol $C$ denotes the corresponding cost functions
\cite{footnote1}. This general criterion is a {\it necessary and
sufficient} condition for the possibility of implementing the
corresponding POVM. The sufficiency here is due to the
close-to-classical character of the totally dephased output states
which are directly linked to the click patterns of the
implementation. In the case of only partially dephased states
corresponding to a conditional-dynamics protocol, the statement in
Eq.~(\ref{generalprinciple}) is no longer sufficient but only
necessary for the implementability of the POVM. Similarly, if the
POVM is not a unique solution to the optimization problem, the
condition in Eq.~(\ref{generalprinciple}) is only necessary.

In general, it will be highly non-trivial to find the quantum
mechanical optimum of the corresponding cost functions. In many
cases, neither for pure states, as typically given before the
dephasing, nor, in particular, for mixed states, as obtained after
dephasing, a closed expression for the optimum exists.

However, for instance, for the non-projective POVM that is an
optimal solution to the unambiguous discrimination of two pure
non-orthogonal states, the corresponding cost function is the
failure probability and its optimum/minimum before dephasing is
simply the overlap (fidelity) of the states. For the mixed states
after dephasing, in this case, at least a lower bound for the cost
function can also be given in terms of the fidelity of the states.
It is then possible to derive a relatively simple set of necessary
conditions for the implementability of the corresponding POVM.
Later we will discuss this example in detail. However, before
applying the general principle in Eq.~(\ref{generalprinciple}) to
non-projective POVMs, let us first review how the known criteria
for projection measurements follow from this principle as a simple
special case.

\section{Projection measurements}\label{Secprojec}

Following the approach of the preceding section, given a
projection measurement, we shall consider this measurement as the
optimal solution to the discrimination of orthogonal states. A
suitable cost function for an error-free state discrimination is
the failure probability, i.e., the probability for obtaining an
inconclusive result. Now the optimal strategy in order to
discriminate states within an orthogonal set is to do a projection
measurement on the space spanned by these orthogonal states. This
strategy will always lead to a conclusive error-free result. Since
this implies zero cost for discriminating orthogonal states,
$C_{\rm quantum\; mechanics}^{\rm optimal} = 0$, a linear-optics
implementation of exact state discrimination means that $C_{\rm
linear\; optics,\; dephasing}^{\rm optimal}\stackrel{!}{=}0$
according to Eq.~(\ref{generalprinciple}).

In order to discriminate any two pure orthogonal states from the
projection measurement basis, the quantum mechanically
optimal/minimal failure probability is given by the overlap of the
states to be discriminated. Expressing the overlap in terms of the
fidelity, $F(\hat\rho_1,\hat\rho_2)\equiv\left({\rm
Tr}\sqrt{\sqrt{\hat\rho_1}\hat\rho_2\sqrt{\hat\rho_1}}\right)^2$,
for two pure orthogonal signal states, $+$ and $-$, of course, we
have $F(\hat\rho_{+},\hat\rho_{-}) = 0$. Hence after dephasing,
the minimal failure probability must not become nonzero, in order
to satisfy our principle in Eq.~(\ref{generalprinciple}). Since in
any mixed-state discrimination scheme, the squared failure
probability is lower bounded by the fidelity of the mixed states
\cite{recentPhilippe}, the condition for implementing the exact
state discrimination becomes $F(\hat\rho_{+,H}',\hat\rho_{-,H}')
\stackrel{!}{=} 0$.
Thus, we have ${\rm Tr}(\hat\rho_{+,H}'\,\hat\rho_{-,H}') = 0$,
since always $0\leq{\rm Tr}(\hat\rho_1\,\hat\rho_2)\leq
F(\hat\rho_1,\hat\rho_2)$. Using the dephasing integral from
Eq.~(\ref{dephasingmechanism}), one can then derive a hierarchy of
simple conditions for the exact discrimination of two or even more
states \cite{PvLNorbert}. These conditions are necessary and
sufficient for the possibility of exactly implementing the
corresponding projection measurement.

For a two-dimensional projection measurement, corresponding to the
discrimination of two orthogonal states $|\chi_{+}\rangle$ and
$|\chi_{-}\rangle$, the necessary and sufficient conditions for an
exact implementation via linear optics and, for instance, photon
counting, are given by \cite{PvLNorbert},
\begin{eqnarray}\label{orthogonalhierarchy2}
\langle\chi_{+}|\hat c^\dagger_j\hat c_j |\chi_{-}\rangle &=& 0\,,
\quad\forall j\;,\\
\langle\chi_{+}|\hat c^\dagger_j\hat c^\dagger_{j'} \hat c_j\hat
c_{j'} |\chi_{-}\rangle &=& 0\,,
\quad\forall j,j'\;, \nonumber\\
\langle\chi_{+}|\hat c^\dagger_j\hat c^\dagger_{j'} \hat
c^\dagger_{j''}\hat c_j\hat c_{j'} \hat c_{j''} |\chi_{-}\rangle
&=& 0\,,
\quad\forall j,j',j''\;, \nonumber \\
\quad\quad\quad\vdots\quad\quad\quad &=&\quad\quad\quad \vdots
\nonumber
\end{eqnarray}
Here, the mode operators $\hat c_j = \hat U^\dagger \hat a_j \hat
U = \sum_i U_{ji} \hat a_i$ are those corresponding to the output
modes of the linear-optics circuit. In the remainder of this
section, we will add some new and useful observations to the
results of Ref.~\cite{PvLNorbert} on projection measurements.

Assuming signal states with a {\it fixed number of photons} (say
$N$ photons), there is an obvious interpretation for the highest
order conditions (i.e., the $N$th order conditions), because for
these we have
\begin{eqnarray}\label{highestorder}
\langle\chi_{+}|\hat c^\dagger_j\hat c^\dagger_{j'} \hat
c^\dagger_{j''}\cdots\hat c_j\hat c_{j'} \hat c_{j''}\cdots
|\chi_{-}\rangle =\quad\quad\quad&&
\nonumber\\
\langle\chi_{+,H}|\hat a^\dagger_j\hat a^\dagger_{j'} \hat
a^\dagger_{j''}\cdots\hat a_j\hat a_{j'} \hat a_{j''}\cdots
|\chi_{-,H}\rangle \propto\quad\quad\quad&&
\nonumber\\
\Psi^*(j,j',j'',...|+)\times \Psi(j,j',j'',...|-) \,,&&
\end{eqnarray}
where $\Psi(j,j',j'',...|\pm)$ is the probability amplitude for
detecting a photon in mode $j$ and another photon in mode $j'$,
etc., when the input was the $+$ or $-$ state. Thus
$\Psi(j,j',j'',...|\pm)$ represents the probability amplitude for
any possible pattern to be detected at the output.

Now it becomes clear why any highest order must vanish for exact
state discrimination. Only those patterns that do not occur at all
and the successful patterns that can be triggered only by one of
the two states lead to $\Psi^*(j,j',j'',...|+)\times
\Psi(j,j',j'',...|-)=0$. In contrast, for any failure pattern, the
product of the probability amplitudes becomes nonzero,
$\Psi^*(j,j',j'',...|+)\times \Psi(j,j',j'',...|-)\neq 0$. As a
result, the highest order conditions {\it alone} are {\it
necessary and sufficient} for exact state discrimination.
Fulfilling all the highest order conditions then implies that all
lower order conditions are satisfied as well. However, note that
the converse does not hold. The lower order conditions are only
necessary, but not sufficient for exact state discrimination.
Thus, if the lower order conditions are satisfied, the highest
order conditions may well be violated. As the lower orders are
easier to calculate than the higher orders, one would normally
start by computing the lowest orders. In order to rule out the
possibility of exact state discrimination, it is then sufficient
to find a violation of any lower order condition (``no-go"
statement). However, for a ``go" statement, the lower orders alone
do not suffice. In this case, for verifying that exact state
discrimination is possible, one has to either calculate the higher
orders as well or directly check a possible solution inferred from
the lower orders. All these observations also indicate that for
unambiguously discriminating two nonorthogonal signal states of
fixed photon number, there must be at least one highest order
condition that is violated (corresponding to the existence of at
least one failure pattern and hence a nonzero failure
probability).

Let us now consider non-projective POVMs including the optimal
unambiguous discrimination of nonorthogonal states via linear
optics.

\section{Non-projective POVMs}\label{Secnonprojec}

Our goal is now, similar to the criteria for projection
measurements, to derive relatively simple conditions for the
implementation of a given non-projective POVM. Our approach shall
be based upon the general principle expressed in
Eq.~(\ref{generalprinciple}).

We have seen already that there are state estimation problems with
trivial optimal POVM solutions. For instance, discriminating
orthogonal states optimally means to perform the corresponding
projection measurement. A very natural way to optimally
discriminate quantum states drawn from a set of linearly
independent states is to perform a POVM that minimizes the
probabiltity of identifying the wrong states. This so-called
minimum error discrimination (MED) can always be described by a
projection measurement onto a suitably chosen basis in the signal
Hilbert space \cite{Helstrom}. Therefore, in order to decide
whether for a given set of quantum states MED can be implemented
via linear optics, we can also directly apply the conditions for
projection measurements. An example for this is the MED of two
symmetric coherent states $|\pm\alpha\rangle$ which cannot be
accomplished via non-asymptotic linear-optics schemes
\cite{Takeoka}.

Another trivial example is the optimal estimation of an unknown
qubit state. In this case, the optimal mean fidelity $\bar F_{\rm
quantum\; mechanics}^{\rm optimal} = 2/3$ \cite{footnote1} can be
attained by randomly choosing an arbitrary qubit basis, measuring
in this basis, and estimating the state via the basis vector that
corresponds to the outcome of the measurement. Thus, trivially,
the optimal estimation of a completely unknown qubit state
$\alpha|\bar 0\rangle + \beta|\bar 1\rangle$ in photonic dual-rail
encoding, $|\bar 0\rangle\equiv |10\rangle$, $|\bar 1\rangle\equiv
|01\rangle$, can be implemented by directly detecting the photons
in the two modes. In fact, in order to satisfy our general
principle in Eq.~(\ref{generalprinciple}), we need to fulfil $1 -
\bar F_{\rm linear\; optics, dephasing}^{\rm optimal}=1 - \bar
F_{\rm quantum\; mechanics}^{\rm optimal}= 1/3$; this can be
accomplished by directly dephasing the input state
\cite{footnote2}.

An example that leads to highly non-trivial POVM solutions is the
calculation of the accessible information in quantum
communication, involving an extremely difficult optimization
problem. Although our general principle expressed in
Eq.~(\ref{generalprinciple}) applies to this problem as well, here
we are not going to attempt to treat a linear-optics
implementation of the accessible information gain.

By contrast, a relatively simple optimization leads to the optimal
unambiguous discrimination of quantum states, i.e., a scheme that
either identifies the signal state correctly or it yields an
inconclusive result with the smallest probability allowed by
quantum theory. In this case, the cost function is the probability
for obtaining an inconclusive result. Now it has been shown that
in general, this failure probability squared has a lower bound
determined by the fidelity of the signal states, ${\rm Prob}_{\rm
fail}^2\geq F$ \cite{recentPhilippe}. For two pure non-orthogonal
signal states, the minimal failure probability squared exactly
coincides with the overlap (fidelity) of the two states (assuming
equal a priori probabilities \cite{IDP1,IDP2,IDP3}). Thus, as for
implementing optimal unambiguous state discrimination (USD), we
can directly apply our general principle to the fidelities of the
states before and after dephasing. The corresponding optimal POVM
solution is a non-trivial non-projective POVM, consisting of two
POVM elements for the correct identification of the states and one
that describes the inconclusive result
\cite{footnote3,TonyPhysLett,Tonyprivate}. Let us now consider the
question whether this optimal USD of {\it two pure} non-orthogonal
states can be implemented with linear optics.

\subsection{Optimal unambiguous state discrimination}

The optimal unambiguous state discrimination (USD) of two pure
non-orthogonal states
\begin{eqnarray}\label{USDinputstateslogicalbasis}
|\chi_+\rangle &=& \alpha |\bar 0\rangle + \beta |\bar 1\rangle\;,
\nonumber\\
|\chi_-\rangle &=& \alpha |\bar 0\rangle - \beta |\bar 1\rangle\;,
\end{eqnarray}
where $\alpha>\beta$ are assumed to be real and $\{|\bar
0\rangle,|\bar 1\rangle\}$ are two basis states, corresponds to a
projection onto the orthogonal set (see also
App.~\ref{onephoton}),
\begin{equation}\label{Naimark}
|w_\mu\rangle = |u_\mu\rangle + |N_\mu\rangle \;,
\end{equation}
in an extended Hilbert space. Here, the $\{|u_\mu\rangle\}$ are
state vectors in a Hilbert space $\cal K$ such that
\begin{equation}\label{POVMoperators}
\hat E_\mu = |u_\mu\rangle\langle u_\mu |
\end{equation}
are the POVM operators of a three-valued POVM, $\mu=1,2,3$, with
$\sum_\mu\hat E_\mu=\mbox{1$\!\!${\large 1}}$. The vectors
$\{|N_\mu\rangle\}$ are defined in the complementary space $\cal
K^\bot$ orthogonal to $\cal K$, with the total Hilbert space $\cal
H = \cal K \oplus \cal K^\bot$. For the optimal USD, one can show
that
\begin{eqnarray}\label{setUSD}
|u_{1/2}\rangle &=&\frac{1}{\sqrt{2}}\left( \frac{\beta}{\alpha}\,
|\bar 0\rangle \pm |\bar 1\rangle
\right) \;,\nonumber\\
|N_{1/2}\rangle &=&
\frac{1}{\sqrt{2}}\sqrt{1-\frac{\beta^2}{\alpha^2}}\,
|\bar 2\rangle \;,\nonumber\\
|u_3\rangle &=&\sqrt{1-\frac{\beta^2}{\alpha^2}}\, |\bar 0\rangle
\;,\quad |N_3\rangle = -\frac{\beta}{\alpha}\,|\bar 2\rangle \;,
\end{eqnarray}
and $\langle\bar 2|\bar 0\rangle=\langle\bar 2|\bar 1\rangle =0$.
The first two POVM elements ($\mu = 1,2$) here refer to the two
signal states, whereas the third POVM element ($\mu = 3$)
corresponds to the inconclusive result. To make the discrimination
unambiguous, we have indeed ${\rm Tr}(\hat E_1
|\chi_-\rangle\langle \chi_-|)= {\rm Tr}(\hat E_2
|\chi_+\rangle\langle \chi_+|)=0$ with $\hat E_\mu$ from
Eq.~(\ref{POVMoperators}) and Eq.~(\ref{setUSD}). To make it
optimal, we have
\begin{eqnarray}\label{USDexample}
{\rm Prob}_{\rm succ} &=&{\rm Tr}(\hat E_1 |\chi_+\rangle\langle
\chi_+|)/2+ {\rm Tr}(\hat E_2 |\chi_-\rangle\langle \chi_-|)/2
\nonumber\\
&=&1-{\rm Prob}_{\rm fail}
\nonumber\\
&=&1-{\rm Tr}(\hat E_3 |\chi_+\rangle\langle \chi_+|)/2- {\rm
Tr}(\hat E_3 |\chi_-\rangle\langle \chi_-|)/2
\nonumber\\
&=&1-|\langle\chi_+|\chi_-\rangle|=1-(\alpha^2-\beta^2)
=2\,\beta^2\;.\nonumber\\
\end{eqnarray}

As for the linear-optical implementation, using one-photon signal
states and multiple-rail encoding, $|\bar 0\rangle\equiv
|100\rangle$, $|\bar 1\rangle\equiv  |010\rangle$, $|\bar
2\rangle\equiv |001\rangle$, one can directly implement the
corresponding POVM for the optimal USD, as described in
App.~\ref{onephoton} for general single-photon based POVMs
[Eq.~(\ref{singleph1}) and Eq.~(\ref{singleph2})]. In this case,
the output states after the linear-optics circuit, $|100\rangle$,
$|010\rangle$, and $|001\rangle$, uniquely refer to one of the
three orthogonal states $|w_\mu\rangle$, and hence identify the
signal states $|\chi_+\rangle$ and $|\chi_-\rangle$ with the best
possible probability. However, in general, for arbitrary signal
states, it turns out to be very hard to decide whether the optimal
USD can be implemented, because of the infinite number of possible
Naimark extensions. In the following, we will investigate the
optimal USD of two pure states independent of the Naimark
extension, using the general principle introduced in the preceding
sections and expressed in Eq.~(\ref{generalprinciple}).

A suitable cost function for the USD of two pure non-orthogonal
states is the failure probability. When optimized over all
possible POVMs, the minimal failure probability corresponds to the
overlap of the states. Thus, according to
Eq.~(\ref{generalprinciple}) and since after dephasing the
mixed-state USD failure probability is bounded from below by the
fidelity \cite{recentPhilippe}, we obtain the condition
\begin{eqnarray}\label{generalrule}
F(\hat\rho_{+,H}',\hat\rho_{-,H}')\stackrel{!}{=}
F(\hat\rho_{+},\hat\rho_{-})\,,
\end{eqnarray}
where $F(\hat\rho_{+},\hat\rho_{-})$ is the fidelity of the input
states and $F(\hat\rho_{+,H}',\hat\rho_{-,H}')$ is the fidelity
after linear optics and dephasing. Note that the fidelity of the
dephased density matrices only yields a lower bound on the failure
probability and the optimal failure probability may well exceed
this bound. Thus, even for a fixed array of linear optics
\cite{footnote4}, described by totally dephased density matrices,
the criterion in Eq.~(\ref{generalrule}) is, in general, only a
{\it necessary condition} for optimal USD. As a result, for the
optimal USD of two pure states via linear optics and subsequent
photon counting (of all modes after static linear optics or only
one first mode in a conditional-dynamics scheme), we have the
following rule: the linear-optics circuit must be chosen such that
{\it the overlap of the two states in terms of the fidelity is the
same before and after dephasing.} This statement, as expressed by
Eq.~(\ref{generalrule}), extends the exact discrimination of
orthogonal states to the more general scenario for optimal
discrimination of nonorthogonal states. Whether linear optics or,
more generally, linear transformations including multi-mode
squeezing (corresponding to arbitrary quadratic interactions) are
sufficient to implement optimal USD depends on the ability of
these tools to obey the above rule. When focusing on the special
case of USD, a more direct derivation of the fidelity criterion in
Eq.~(\ref{generalrule}) is possible and given in
App.~\ref{rigorous}. Let us now examine the statement in
Eq.~(\ref{generalrule}) in more detail for a fixed array of linear
optics.

\subsection{Optimal USD via a fixed linear network}

For a fixed array of linear optics, all output modes will be
detected at once. Therefore, Eq.~(\ref{generalrule}) refers to
totally dephased density matrices. In order to check the criterion
in Eq.~(\ref{generalrule}), we find that the fidelity before and
after linear optics becomes
\begin{eqnarray}\label{fidbeforeafter}
F(\hat\rho_{+},\hat\rho_{-}) = F(\hat\rho_{+,H},\hat\rho_{-,H}) =
\sum_{m,n} \alpha_m^*\alpha_n\beta_m\beta_n^*\,,
\end{eqnarray}
because after the linear-optics transformation, the output states
will always take on the following form,
\begin{eqnarray}\label{nonorthogstatesafterlin}
|\chi_{+,H}\rangle &=& \sum_{k}\alpha_{k} |\{k\}\,\rangle
+\sum_{m}\alpha_{m} |\{m\}\,\rangle , \nonumber\\
|\chi_{-,H}\rangle &=& \sum_{l}\beta_{l} |\{l\}\,\rangle +
\sum_{m}\beta_{m} |\{m\}\,\rangle \,,
\end{eqnarray}
where the coefficients depend on the linear-optics circuit chosen
in a particular implementation. The indices $k$ and $l$ denote
photon number patterns, i.e., $N$-mode Fock states, that
exclusively occur in the expansion of $|\chi_{+,H}\rangle$ and
$|\chi_{-,H}\rangle$, respectively. Hence these patterns
unambiguously refer to the $+$ state or to the $-$ state. However,
because of the finite overlap of the input states, we must include
patterns that occur in the expansion of both states. These
ambiguous patterns are denoted by the index $m$. In general, the
amplitudes of the ambiguous $N$-mode Fock states in the
expansions, and hence the probabilities for the corresponding
patterns to be detected, may be different for the $+$ and the $-$
state.

After dephasing, the output states take on the following form
\begin{eqnarray}\label{dephasedstatesafterlin}
\hat\rho_{+,H}' &=& \sum_{k}{\rm P}^+_{k}
|\{k\}\,\rangle\langle\{k\}|
+\sum_{m}{\rm P}^+_{m} |\{m\}\,\rangle\langle \{m\}| , \nonumber\\
\hat\rho_{-,H}' &=& \sum_{l}{\rm P}^-_{l}
|\{l\}\,\rangle\langle\{l\}| + \sum_{m}{\rm P}^-_{m}
|\{m\}\,\rangle\langle\{m\}| ,\nonumber\\
\end{eqnarray}
corresponding to a dephasing of the states in
Eq.~(\ref{nonorthogstatesafterlin}) with the probabilities given
by ${\rm P}^+_{k} = |\alpha_k|^2$, ${\rm P}^-_{l} = |\beta_l|^2$,
${\rm P}^+_{m} = |\alpha_m|^2$, and ${\rm P}^-_{m} = |\beta_m|^2$.

The fidelity after linear optics and dephasing is now given by
\begin{eqnarray}\label{fidafterlinoptanddephase}
F(\hat\rho_{+,H}',\hat\rho_{-,H}') &=& \left({\rm
Tr}\sqrt{\hat\rho_{+,H}'\,\hat\rho_{-,H}'}\right)^2 \nonumber\\
&=& \left(\sum_m\sqrt{{\rm P}^+_{m}{\rm P}^-_{m}}\right)^2.
\end{eqnarray}
Thus, the fidelity criterion from Eq.~(\ref{generalrule}) can be
expressed by
\begin{eqnarray}
\sum_{m,n}\sqrt{{\rm P}^+_{m}{\rm P}^+_{n}{\rm P}^-_{m}{\rm
P}^-_{n}} \stackrel{!}{=} \sum_{m,n}
\alpha_m^*\alpha_n\beta_m\beta_n^*\,,
\end{eqnarray}
using Eq.~(\ref{fidafterlinoptanddephase}) and
Eq.~(\ref{fidbeforeafter}). This, however, implies that
\begin{eqnarray}\label{condition1}
\sum_{m,n} |\alpha_m||\alpha_n||\beta_m||\beta_n|
&\stackrel{!}{=}&
\sum_{m,n} |\alpha_m||\alpha_n||\beta_m||\beta_n|\nonumber\\
&\times& e^{i(\phi_m^--\phi_n^-+\phi_n^+-\phi_m^+)}\,,
\end{eqnarray}
where $\alpha_m = |\alpha_m|e^{i\phi_m^+}$ and $\beta_m =
|\beta_m|e^{i\phi_m^-}$, etc. The only possible way to satisfy
Eq.~(\ref{condition1}) is for
$e^{i(\phi_m^--\phi_n^-+\phi_n^+-\phi_m^+)} = 1$, $\forall m,n$.
Thus, we have $\phi_m^- - \phi_m^+ = \phi$, $\forall m$. A direct
consequence of this result is that the overlap of the input states
can be written as
\begin{eqnarray}\label{newoverlap}
|\langle \chi_{+}|\chi_{-}\rangle | &=& |\langle
\chi_{+,H}|\chi_{-,H}\rangle | \nonumber\\
&=& \left| \sum_m \alpha_m^* \beta_m \right| = \sum_m |\alpha_m|
|\beta_m|.
\end{eqnarray}
Let us now look at the first-order expression from the conditions
in Eq.~(\ref{orthogonalhierarchy2}). We obtain
\begin{eqnarray}\label{nonorthogonalstate1storderfromfid}
\langle\chi_{+}|\hat c^\dagger_j \hat c_j|\chi_{-}\rangle &=&
\langle\chi_{+,H}|\hat a^\dagger_j \hat a_j|\chi_{-,H}\rangle
\\
&=& e^{i\phi} \sum_m |\alpha_m||\beta_m|  \langle \{m\}\,|\hat
a^\dagger_j \hat a_j|\{m\}\,\rangle \,,\nonumber
\end{eqnarray}
because annihilating a photon in the $j$th mode of both states
only leads to nonzero contributions from coinciding patterns.
Using Eq.~(\ref{newoverlap}) and
Eq.~(\ref{nonorthogonalstate1storderfromfid}), there are two
observations we can make. First, the modulus of any first order
expression $\langle\chi_{+}|\hat c^\dagger_j \hat
c_j|\chi_{-}\rangle$ is bounded from above such that
\begin{eqnarray}\label{nonorthogonalstatenew1stordermodulus}
\left|\langle\chi_{+}|\hat c^\dagger_j \hat
c_j|\chi_{-}\rangle\right| \leq
N\,\left|\langle\chi_{+}|\chi_{-}\rangle\right|\,,\forall j\,,
\end{eqnarray}
where $N$ is the maximum photon number in the states. In addition,
we have
\begin{eqnarray}\label{nonorthogonalstatenew1storder}
\frac{\langle\chi_{+}|\hat c^\dagger_j \hat c_j|\chi_{-}\rangle}{
\left|\langle\chi_{+}|\hat c^\dagger_j \hat
c_j|\chi_{-}\rangle\right|} = \frac{\langle\chi_{+}|\hat
c^\dagger_{j'} \hat c_{j'}|\chi_{-}\rangle}{
\left|\langle\chi_{+}|\hat c^\dagger_{j'} \hat
c_{j'}|\chi_{-}\rangle\right|}\,,\forall j,j'\,,
\end{eqnarray}
provided that $\langle\chi_{+}|\hat c^\dagger_j \hat
c_j|\chi_{-}\rangle$ and $\langle\chi_{+}|\hat c^\dagger_{j'} \hat
c_{j'}|\chi_{-}\rangle$ are both nonzero. Similarly, for the
second-order expressions, we obtain
\begin{eqnarray}\label{nonorthogonalstate2ndorder}
\langle\chi_{+}|\hat c^\dagger_j\hat c^\dagger_{j'} \hat c_j\hat
c_{j'}|\chi_{-}\rangle &=& \langle\chi_{+,H}|\hat a^\dagger_j\hat
a^\dagger_{j'} \hat a_j\hat a_{j'}|\chi_{-,H}\rangle
\\
&=& e^{i\phi} \sum_m |\alpha_m||\beta_m| \nonumber\\
&&\quad\times \langle \{m\}\,|\hat a^\dagger_j \hat
a^\dagger_{j'}\hat a_j\hat a_{j'}|\{m\}\,\rangle .\nonumber
\end{eqnarray}
This leads to
\begin{eqnarray}\label{nonorthogonalstatenew2ndordermodulus}
\left|\langle\chi_{+}|\hat c^\dagger_j\hat c^\dagger_{j'} \hat
c_j\hat c_{j'}|\chi_{-}\rangle\right| \leq
N(N-1)\,\left|\langle\chi_{+}|\chi_{-}\rangle\right|\,,\forall
j,j',\nonumber\\
\end{eqnarray}
and
\begin{eqnarray}\label{nonorthogonalstatenew2ndorder}
\frac{\langle\chi_{+}|\hat c^\dagger_j\hat c^\dagger_{i} \hat
c_j\hat c_{i}|\chi_{-}\rangle}{ \left|\langle\chi_{+}|\hat
c^\dagger_j\hat c^\dagger_{i} \hat c_j\hat
c_{i}|\chi_{-}\rangle\right|} = \frac{\langle\chi_{+}|\hat
c^\dagger_{j'}\hat c^\dagger_{i'} \hat c_{j'}\hat
c_{i'}|\chi_{-}\rangle}{ \left|\langle\chi_{+}|\hat
c^\dagger_{j'}\hat c^\dagger_{i'} \hat c_{j'}\hat
c_{i'}|\chi_{-}\rangle\right|}\,,\forall j,i,j',i'\,,\nonumber\\
\end{eqnarray}
provided that $\langle\chi_{+}|\hat c^\dagger_j\hat c^\dagger_{i}
\hat c_j\hat c_{i}|\chi_{-}\rangle$ and $\langle\chi_{+}|\hat
c^\dagger_{j'}\hat c^\dagger_{i'} \hat c_{j'}\hat
c_{i'}|\chi_{-}\rangle$ are both nonzero. Moreover, for all
non-vanishing expressions, also the phases of different orders
must coincide. As a result, we have proven the following theorem
for the implementability of optimal USD of two pure nonorthogonal
states with linear optics.

{\bf Theorem:} it is a necessary (but, in general, not sufficient)
criterion for the possibility of implementing the optimal USD of
two pure nonorthogonal states $|\chi_\pm\rangle$ via static linear
optics and photon counting that the hierarchies of conditions,
\begin{eqnarray}\label{nonorthogonalstatenewhierarchy}
&&\frac{\langle\chi_{+}|\chi_{-}\rangle}{
\left|\langle\chi_{+}|\chi_{-}\rangle\right|} \nonumber\\
&=& \frac{\langle\chi_{+}|\hat c^\dagger_j \hat
c_j|\chi_{-}\rangle}{ \left|\langle\chi_{+}|\hat c^\dagger_j \hat
c_j|\chi_{-}\rangle\right|} = \frac{\langle\chi_{+}|\hat
c^\dagger_{j'} \hat c_{j'}|\chi_{-}\rangle}{
\left|\langle\chi_{+}|\hat c^\dagger_{j'} \hat
c_{j'}|\chi_{-}\rangle\right|} \nonumber\\
&=& \frac{\langle\chi_{+}|\hat c^\dagger_j\hat c^\dagger_{i} \hat
c_j\hat c_{i}|\chi_{-}\rangle}{ \left|\langle\chi_{+}|\hat
c^\dagger_j\hat c^\dagger_{i} \hat c_j\hat
c_{i}|\chi_{-}\rangle\right|} = \frac{\langle\chi_{+}|\hat
c^\dagger_{j'}\hat c^\dagger_{i'} \hat c_{j'}\hat
c_{i'}|\chi_{-}\rangle}{ \left|\langle\chi_{+}|\hat
c^\dagger_{j'}\hat c^\dagger_{i'} \hat c_{j'}\hat
c_{i'}|\chi_{-}\rangle\right|}\,, {\rm etc.}\,,\nonumber\\
&&\quad\quad\quad\quad\quad\quad\forall j,i,j',i', {\rm etc.}\,,
\end{eqnarray}
for any non-vanishing orders, and
\begin{eqnarray}\label{nonorthogonalstatenewhierarchymodulus}
\left|\langle\chi_{+}|\hat c^\dagger_j \hat
c_j|\chi_{-}\rangle\right| \leq
N\,\left|\langle\chi_{+}|\chi_{-}\rangle\right|\,,\forall j\,,
\nonumber\\
\left|\langle\chi_{+}|\hat c^\dagger_j\hat c^\dagger_{j'} \hat
c_j\hat c_{j'}|\chi_{-}\rangle\right| \leq
N(N-1)\,\left|\langle\chi_{+}|\chi_{-}\rangle\right|\,,\forall
j,j',\nonumber\\
\quad\quad\quad\vdots\quad\quad\quad \leq\quad\quad\quad \vdots
\quad\quad\quad\quad\quad\quad\quad\quad\quad\\
\left|\langle\chi_{+}|\hat c^\dagger_j\hat c^\dagger_{j'} \hat
c^\dagger_{j''}\cdots\hat c_j\hat c_{j'} \hat c_{j''}\cdots
|\chi_{-}\rangle\right| \leq \quad\quad\quad\quad\quad\quad\nonumber\\
N!\,\left|\langle\chi_{+}|\chi_{-}\rangle\right|\,,\forall
j,j',j'',...,\nonumber
\end{eqnarray}
are satisfied, where $N$ is the maximum photon number in the
states. For the existence of a linear-optics solution to the POVM
that realizes the optimal USD, output mode operators $\hat c_j$
must be found such that a unitary matrix $U$ can be constructed
with $\hat c_j = \sum_i U_{ji} \hat a_i$ and the hierarchies of
conditions are satisfied for these operators.

Note that for fixed photon number,
the first set of conditions of the theorem [that for the phases in
Eq.~(\ref{nonorthogonalstatenewhierarchy})] implies the second one
[that for the absolute values in
Eq.~(\ref{nonorthogonalstatenewhierarchymodulus})], however, the
converse does not hold. The reason is that in {\it any}
linear-optics scheme, when the input states have a fixed number of
photons $N$, the total sum of a given order satisfies a relation
similar to
\begin{eqnarray}\label{anyscheme}
\langle\chi_{+}|\sum_j\hat c^\dagger_j\hat c_j |\chi_{-}\rangle
&=& N\,\langle\chi_{+}|\chi_{-}\rangle \,.
\end{eqnarray}
Here, the sum of the first-order expressions leads to the total
photon number operator for the output modes, and the relation in
Eq.~(\ref{anyscheme}) follows from the photon number conservation
property of linear optics. Analogous conditions can be found for
the total sum of the higher order expressions. Now for the sum of
the first orders, for instance, according to
Eq.~(\ref{anyscheme}), we obtain
\begin{eqnarray}\label{anyschemeagain}
\left|\sum_j\langle\chi_{+}|\hat c^\dagger_j\hat c_j
|\chi_{-}\rangle\right| =
N\,\left|\langle\chi_{+}|\chi_{-}\rangle\right| \,,
\end{eqnarray}
and, provided the phases of all non-vanishing first orders
coincide as required to obey
Eq.~(\ref{nonorthogonalstatenewhierarchy}),
\begin{eqnarray}\label{anyschemeagain2}
\left|\sum_j\langle\chi_{+}|\hat c^\dagger_j\hat c_j
|\chi_{-}\rangle\right| = \sum_j\left|\langle\chi_{+}|\hat
c^\dagger_j\hat c_j |\chi_{-}\rangle\right| \,.
\end{eqnarray}
The two relations of Eq.~(\ref{anyschemeagain}) and
Eq.~(\ref{anyschemeagain2}) together imply that the first-order
conditions in Eq.~(\ref{nonorthogonalstatenewhierarchymodulus})
are automatically satisfied by any linear-optics transformation
that fulfils the first-order conditions in
Eq.~(\ref{nonorthogonalstatenewhierarchy}).

The hierarchies of conditions in
Eq.~(\ref{nonorthogonalstatenewhierarchy}) and
Eq.~(\ref{nonorthogonalstatenewhierarchymodulus}) are necessary
for optimal USD of two nonorthogonal states. In words, these
criteria mean that for optimal USD, {\it the phases of all
non-vanishing orders must coincide}. For example, any real orders
must have the same sign in optimal USD. In the special case of
orthogonal states, $\langle\chi_{+}|\chi_{-}\rangle = 0$, one can
easily see in Eq.~(\ref{nonorthogonalstatenewhierarchymodulus})
that the hierarchy for exact state discrimination from
Eq.~(\ref{orthogonalhierarchy2}) can be retrieved. An alternative
derivation of the conditions in
Eq.~(\ref{nonorthogonalstatenewhierarchy}) and
Eq.~(\ref{nonorthogonalstatenewhierarchymodulus}), independent of
the fidelity criterion in Eq.~(\ref{generalrule}), is given in
App.~\ref{alternative}.


Let us now look at an example of two nonorthogonal states with
only two photons (which is the simplest nontrivial extension to
the trivial case of one-photon states). We are going to consider
the two-photon toy model state $\alpha |20\rangle \pm \beta
|11\rangle$, where, without loss of generality, $\alpha$ and
$\beta$ are assumed to be real. For the special case of an
orthogonal pair $\alpha = \beta$, it is know that there is no
linear-optics solution for optimally and hence exactly
discriminating these two states, including feedforward and
arbitrary auxiliary states \cite{PvLNorbert}. Here we consider the
nonorthogonal case without auxiliary photons, but arbitrarily many
additional vacuum modes.

Defining $U_{j1}\equiv\nu_1$ and $U_{j2}\equiv\nu_2$ for the
elements of the $j$th row of the unitary matrix in $\hat c_j =
\sum_i U_{ji} \hat a_i$, the first-order expression for mode $j$
then becomes
\begin{eqnarray}\label{toyex1}
\langle\chi_{+}|\hat c^\dagger_j \hat c_j|\chi_{-}\rangle &=&
2\alpha^2|\nu_1|^2 - \beta^2(|\nu_1|^2 + |\nu_2|^2)
\nonumber\\
&&+ \sqrt{2}\alpha\beta (\nu_1\nu_2^* - c.c.)\,.
\end{eqnarray}
Now assuming a {\it fixed array} of linear optics, the conditions
in Eq.~(\ref{nonorthogonalstatenewhierarchy}) are necessary for
optimal USD. In order to satisfy the first-order conditions there
for any modes $j$, $j'$, etc., the expression in
Eq.~(\ref{toyex1}) must become real for any $j$, $j'$, etc.,
because the $0$th order $\langle\chi_{+}|\chi_{-}\rangle =
\alpha^2 - \beta^2$ is real. The same argument applies to the
second-order expressions,
\begin{eqnarray}\label{toyex2}
\langle\chi_{+}|(\hat c^\dagger_j)^2 \hat c_j^2|\chi_{-}\rangle
&=& 2|\nu_1|^2 \big[\alpha^2|\nu_1|^2 - 2\beta^2 |\nu_2|^2
\nonumber\\
&&+ \sqrt{2}\alpha\beta (\nu_1\nu_2^* - c.c.)\big]\,.
\end{eqnarray}
Knowing that all these expressions must be real, let us evaluate
the first-order and second-order conditions for positive and
negative $0$th order $\langle\chi_{+}|\chi_{-}\rangle$, $\alpha^2
> \beta^2$ or $\alpha^2 < \beta^2$, respectively. According to
Eq.~(\ref{nonorthogonalstatenewhierarchy}), we obtain
\begin{eqnarray}\label{toyex3}
\vec\nu\;\vec a\geq 0\,,\quad{\rm and}\quad \vec\nu\;\vec b\geq
0\,,
\end{eqnarray}
for $\alpha^2 > \beta^2$, and
\begin{eqnarray}\label{toyex4}
\vec\nu\;\vec a\leq 0\,,\quad{\rm and}\quad \vec\nu\;\vec b\leq
0\,,
\end{eqnarray}
for $\alpha^2 < \beta^2$, where
\begin{eqnarray}\label{toyex5}
\vec\nu \equiv \left( \begin{array}{c} |\nu_1|^2 \\
|\nu_2|^2
\end{array} \right),\quad
\vec a \equiv \left( \begin{array}{c} 2\alpha^2-\beta^2 \\
-\beta^2
\end{array} \right),\quad
\vec b \equiv \left( \begin{array}{c} \alpha^2 \\
-2 \beta^2
\end{array} \right).\nonumber\\
\end{eqnarray}
In Eq.~(\ref{toyex3}), $\vec\nu\;\vec b\geq 0$ implies that
$\alpha^2\geq 2\beta^2$, because otherwise, for $\alpha^2 <
2\beta^2$, the only way to prevent $\vec\nu\;\vec b$ from becoming
negative is to have $|\nu_1|^2 > |\nu_2|^2$ for any modes $j$,
$j'$, etc., according to
Eq.~(\ref{nonorthogonalstatenewhierarchy}). However, no unitary
matrix can be constructed, where all the elements $j$, $j'$, etc.,
in the first two columns satisfy $|\nu_1|^2 > |\nu_2|^2$.
Similarly, in Eq.~(\ref{toyex4}), $\vec\nu\;\vec a\leq 0$ leads to
$\alpha^2\leq \beta^2$ (which is also simply given by the $0$th
order). Thus, there is a regime, $\beta^2\leq\alpha^2 < 2\beta^2$
(including the orthogonal case $\alpha^2 = \beta^2$), where
optimal USD is impossible for a fixed array of linear optics and
without auxiliary photons.

For $\alpha^2= 2\beta^2$, the optimal solution is a simple $50/50$
beam splitter, $|\nu_1|^2 = |\nu_2|^2 = 1/2$ for modes $j=1,2$. In
this case, in agreement with Eq.~(\ref{toyex3}), we obtain
$\vec\nu\;\vec a = \beta^2 > 0$ and $\vec\nu\;\vec b = 0$. The
orthogonal set of the corresponding von Neumann measurement
becomes
\begin{eqnarray}
|w_{1/2}\rangle&=&\frac{1}{\sqrt{2}} \left[\frac{1}{\sqrt{2}}
(|20\rangle + |02\rangle)\pm|11\rangle\right]\;,
\nonumber\\
|w_3\rangle&=&\frac{1}{\sqrt{2}} (|20\rangle - |02\rangle)\;,
\end{eqnarray}
choosing for the Naimark extension $|\bar 2\rangle\equiv
|02\rangle$ [see
Eqs.~(\ref{USDinputstateslogicalbasis})-(\ref{setUSD})]. A
symmetric beam splitter turns these three states into
$|20\rangle$, $|02\rangle$, and $|11\rangle$, respectively, which
via photon counting uniquely refer to the three different POVM
elements. Thus, optimal USD for $\alpha=\sqrt{2}\beta$ can be
achieved with a simple beam splitter. The two signal states
$\alpha |20\rangle \pm \beta |11\rangle$ are transformed by the
symmetric beam splitter into $\sqrt{\frac{2}{3}}|20\rangle +
\frac{1}{\sqrt{3}}|11\rangle$ and $\sqrt{\frac{2}{3}}|02\rangle +
\frac{1}{\sqrt{3}}|11\rangle$, respectively. Indeed, we have ${\rm
Prob}_{\rm succ}=2/3=2\beta^2$.

\subsection{Optimal USD via conditional dynamics}

We may also apply the general fidelity criterion to a more
sophisticated linear-optics implementation of state
discrimination, namely one that includes conditional dynamics
(feedforward): instead of detecting all output modes after the
linear-optics circuit, one may select only one mode for detection.
After this first measurement, one can then send the conditional
state of the remaining modes through another linear-optics circuit
which depends on the measurement outcome. In the most general
approach, one can include as many feedforward steps as modes are
in the signal states, or even more by adding auxiliary states.

The extension from a static linear-optics scheme to a scheme that
may include conditional dynamics is straightforward for projection
measurements \cite{PvLNorbert}. In this case, simply the subset of
the fixed-array conditions, referring only to a particular mode
operator $\hat c_j$, is necessary for the exact state
discrimination after detecting a first mode $j$. For instance, for
implementing a two-dimensional projection measurement,
corresponding to the discrimination of two orthogonal states
$|\chi_{+}\rangle$ and $|\chi_{-}\rangle$, we have the subset of
the conditions in Eq.~(\ref{orthogonalhierarchy2}),
\begin{equation}
\label{condcond} \langle\chi_+| \left( \hat c^\dagger_j \right) ^n
\left( \hat c_j \right )^n |\chi_-\rangle = 0 \;,
\quad\quad\forall \, n \geq 0 \,.
\end{equation}
These criteria express the necessary requirement for exact state
discimination that the detection of one mode must be either
conclusive or the orthogonality of the signal states must be
preserved in the conditional states of the remaining modes
\cite{PvLNorbert}. The non-existence of some $\hat c_j$ fulfilling
Eq.~(\ref{condcond}) means that as soon as one output mode is
selected and measured, this will make exact discrimination of the
states impossible.

For the general case, including non-projective POVMs, the
extension from static linear optics to conditional dynamics is
slightly more subtle. A complete derivation of the conditions for
implementing optimal USD via linear optics and feedforward can be
found in App.~\ref{conditionaldynamics}. The resulting conditions
necessary for optimal USD when detecting a first mode $j$ are
again simply the subset of the fixed-array conditions in
Eq.~(\ref{nonorthogonalstatenewhierarchy}) and
Eq.~(\ref{nonorthogonalstatenewhierarchymodulus}) referring to
this one mode; thus, we obtain
\begin{eqnarray}\label{nonorthogonalstatenewhierarchyconddyn}
&&\frac{\langle\chi_{+}|\chi_{-}\rangle}{
\left|\langle\chi_{+}|\chi_{-}\rangle\right|} \nonumber\\
&=& \frac{\langle\chi_{+}|\hat c^\dagger_j \hat
c_j|\chi_{-}\rangle}{ \left|\langle\chi_{+}|\hat c^\dagger_j \hat
c_j|\chi_{-}\rangle\right|} = \frac{\langle\chi_{+}|(\hat
c^\dagger_j)^2 \hat c_j^2|\chi_{-}\rangle}{
\left|\langle\chi_{+}|(\hat c^\dagger_j)^2 \hat
c_j^2|\chi_{-}\rangle\right|}\nonumber\\
&=& \cdots = \frac{\langle\chi_{+}|(\hat c^\dagger_j)^n \hat
c_j^n|\chi_{-}\rangle}{ \left|\langle\chi_{+}|(\hat c^\dagger_j)^n
\hat c_j^n|\chi_{-}\rangle\right|} \,,{\rm etc.}\,,
\end{eqnarray}
for any non-vanishing orders, and
\begin{eqnarray}\label{nonorthogonalstatenewhierarchymodulusconddyn}
\left|\langle\chi_{+}|\hat c^\dagger_j \hat
c_j|\chi_{-}\rangle\right| \leq
N\,\left|\langle\chi_{+}|\chi_{-}\rangle\right|\,,
\nonumber\\
\left|\langle\chi_{+}|(\hat c^\dagger_j)^2 \hat
c_j^2|\chi_{-}\rangle\right|\leq
N(N-1)\,\left|\langle\chi_{+}|\chi_{-}\rangle\right|\,,{\rm etc.}
\end{eqnarray}
These conditions are also a direct consequence of the optimal-USD
fidelity criterion in Eq.~(\ref{generalrule}). However, this time,
only partially dephased density operators, corresponding to the
detection of only one mode, must be considered. These partially
dephased density matrices are, in general, no longer diagonal in
the Fock basis (for details, see App.~\ref{conditionaldynamics}).

In the next section, we examine whether our new set of conditions
enables us to make general statements about the use of auxiliary
photons for the optimal USD of two nonorthogonal states.

\subsection{Auxiliary photons for optimal USD}

Let us consider the following question: can the use of an
auxiliary state make optimal USD via linear optics possible when
it is impossible without an ancilla state. In this case, the input
states to be discriminated become
$|\chi_\pm\rangle=|s_\pm\rangle\otimes|\psi_{\rm aux}\rangle$,
where $|s_\pm\rangle$ represent the signal states and $|\psi_{\rm
aux}\rangle$ is the auxiliary state. The auxiliary state contains
optical modes in addition to the signal modes, and these extra
modes may be occupied by additional photons. For the special case
of projective POVMs, it has been shown already, using the criteria
for projection measurements, that if the orthogonal signal states
contain a fixed number of photons, adding an ancilla state
(including extra photons or not) never helps \cite{PvLNorbert}.
This can be seen by splitting the input modes into a set of signal
and a set of auxiliary modes, thus decomposing the output mode
operator $\hat c_j = \sum_i U_{ji} \hat a_i$ into two
corresponding parts as (dropping the index $j$) $\hat c = b_{\rm
s}\hat c_{\rm s}+ b_{\rm aux}\hat c_{\rm aux}$, with real
coefficients $b_{\rm s}$ and $b_{\rm aux}$, where $\hat c_{\rm s}$
acts only upon the signal modes and $\hat c_{\rm aux}$ only on the
auxiliary modes. Using these output mode operators and assuming
orthogonal signal states with fixed photon number, the criteria
for exact state discrimination are the same with or without
arbitrary ancilla states \cite{PvLNorbert}.

Now for the case of the non-projective POVM for optimal USD of two
pure nonorthogonal states, the same approach as described in the
preceding paragraph will not enable us to make a general
statement. Inserting the output mode operator $\hat c = b_{\rm
s}\hat c_{\rm s}+ b_{\rm aux}\hat c_{\rm aux}$ into, for example,
the first-order expression $\langle\chi_+|\hat c^{\dagger}\hat
c|\chi_-\rangle$ yields a result that, in general, for $\langle
s_+|s_-\rangle\neq 0$ (with either a fixed or an undetermined
photon number), still depends on the auxiliary state.

In general, we cannot rule out the possibility that adding an
ancilla helps to satisfy the conditions for optimal USD when they
cannot be fulfilled without ancilla. Moreover, adding only
auxiliary vacuum modes without extra photons might also be useful
and necessary in order to build up a unitary matrix for the mode
operators $\hat c_j = \sum_i U_{ji} \hat a_i$. In fact, in the
one-photon example discussed after Eq.~(\ref{USDexample}), adding
an extra vacuum mode is essential in order to extend the
two-dimensional signal Hilbert space to an at least
three-dimensional space required for the POVM and to construct the
corresponding unitary matrix.


\begin{figure}[tb]
\epsfxsize=1.2in \epsfbox[220 300 400 560]{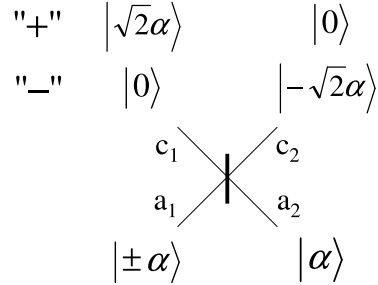}
\caption{\label{fig2} Implementing the optimal unambiguous
discrimination of two symmetric coherent states via a simple
$50/50$ beam splitter and an auxiliary coherent state of the same
amplitude.}
\end{figure}

There are also known examples, where adding extra photons makes
the optimal USD of the two signal states via linear optics
possible. One such example for the case of infinite-dimensional
signal and auxiliary states, both with an {\it undetermined and
unbounded photon number}, is the optimal USD of so-called binary
coherent states. In this case, the optimal USD can be easily
achieved using a $50/50$ beam splitter and an ancilla coherent
state (see Fig.~\ref{fig2}). In our notation, one has
$|s_\pm\rangle \equiv|\pm\alpha\rangle$ ($\alpha$ assumed to be
real), $|\psi_{\rm aux}\rangle\equiv|\alpha\rangle$, and
$|\chi_\pm\rangle=|s_\pm\rangle\otimes|\psi_{\rm aux}\rangle$.
This two-mode state is now transformed by the $50/50$ beam
splitter into
\begin{eqnarray}\label{bincoherentstates}
|\chi_{+,H}\rangle &=& |\sqrt{2}\alpha\rangle\otimes
|0\rangle , \nonumber\\
|\chi_{-,H}\rangle &=& |0\rangle\otimes
|-\sqrt{2}\alpha\rangle
\,.
\end{eqnarray}
For these states, a detector click in mode 1 can only be triggered
by the $+$ state, whereas a click in mode 2 unambiguously refers
to the $-$ state. However, there are inconclusive ``events"
corresponding to the two-mode vacuum state, $|\psi_{\rm
inconcl}\rangle = e^{-\alpha^2}\,|00\rangle$, using
Eq.~(\ref{nonorthogstatesafterlin3dim}) from
App.~\ref{alternative} with $\phi = 0$. Since the failure
probability is then given by ${\rm Prob}_{\rm fail}^{\rm lin.opt.}
= (e^{-2\alpha^2} + e^{-2\alpha^2})/2 = e^{-2\alpha^2} = \langle
+\alpha|-\alpha\rangle$, this scheme turns out to be optimal.
Thus, we expect that the corresponding solution satisfies our
criteria for optimal USD. For a particular mode $j$, again using
$U_{j1}\equiv\nu_1$ and $U_{j2}\equiv\nu_2$ for the elements of
the $j$th row of the unitary matrix in $\hat c_j = \sum_i U_{ji}
\hat a_i$, we obtain the $n$th-order condition,
\begin{eqnarray}
\langle\chi_{+}|(\hat c^\dagger_j)^n \hat c_j^n|\chi_{-}\rangle
&=& \langle +\alpha|-\alpha\rangle \,\Big(|\nu_1|^2\alpha^2
\\
&&- |\nu_2|^2 {\,_2}\langle\psi_{\rm aux}|\hat a_2^\dagger\hat
a_2|\psi_{\rm aux}\rangle_2\Big)^n.\nonumber
\end{eqnarray}
Apparently, for any mode $j=1,2$, any order $n\geq 1$ can be set
to zero by choosing a $50/50$ beam splitter,
$|\nu_1|^2=|\nu_2|^2=1/2$, and the appropriate ancilla state,
$|\psi_{\rm aux}\rangle\equiv|\alpha\rangle$. This solution is
indeed in agreement with the conditions that we derived for
optimal USD. The obvious reason, why all nonzero orders vanish in
this example, is that the only failure pattern here is
$|00\rangle$ which always vanishes upon applying annihilation
operators [see, e.g.,
Eq.~(\ref{nonorthogonalstate1storderfromfid})]. From this
observation follows that also any cross orders for modes 1 and 2
will vanish with the above solution. Let us emphasize again that
in this example, neither the signal nor the auxiliary state
contain a fixed number of photons. For such a scenario, even in
the case of projective POVMs \cite{PvLNorbert}, adding auxiliary
photons may indeed help. However, conversely, even including
non-projective POVMs such as the optimal USD of two pure
nonorthogonal states, we are not aware of any example of a POVM
for signal states with a fixed number of photons where it helps to
add extra photons. Of course, this statement does not apply to
asymptotic schemes \cite{KLM01} for which it is known that
auxiliary photons are, in general, a useful and necessary extra
resource.

Let us finally note that for the optimal USD of more than two
coherent states, symmetrically distributed in phase space, the
optimal USD \cite{TonyPhysLett} cannot be achieved as easily as
for the binary case. However, there are asymptotic linear-optics
solutions including the use of feedforward \cite{vanEnk02}.

\section{Conclusions}

We considered the problem of implementing generalized measurements
(POVMs) with linear optics. Such an implementation may either be
based upon a static array of linear optics or it may include
conditional dynamics (feedforward). Extending our previous results
on projective measurements, we focused, in particular, on
non-projective measurements. Our approach to this problem can be
formulated as a general principle in the following way. We start
by identifying a given POVM as a solution to an optimization
problem for a chosen cost function. The implementation is then
only possible if a linear-optics circuit exists for which the
quantum mechanical optimum is still attainable after dephasing the
corresponding quantum states. As an example for applying this
principle to the problem of implementing a non-projective POVM, we
discussed in detail the optimal unambiguous state discrimination
(USD) of two pure nonorthogonal states. In order to implement the
POVM that realizes the quantum mechanically optimal USD with
linear optics, according to the general principle, the
linear-optics circuit must be chosen such that the overlap of the
states, in terms of the fidelity, is the same before and after
dephasing. This statement extends the exact discrimination of
orthogonal states to the more general scenario for optimal
discrimination of nonorthogonal states. Using the fidelity
criterion, we derived hierarchies of necessary conditions for the
possibility of implementing the optimal USD of two pure
nonorthogonal states via linear optics and photon counting. The
resulting conditions are a generalization of our previous criteria
for projection measurements and the exact discrimination of
orthogonal states.

As for the detection mechanism, here we only studied the case of
photon counting which leads to dephased states diagonal in the
Fock basis. Potential extensions of our results may include
different detection mechanisms such as homodyne detection, as we
discussed previously already in the context of projective
measurements. Moreover, apart from passive linear-optics circuits,
our criteria can also be applied to arbitrary linear mode
transformations, including multi-mode squeezing. When analyzing
those POVMs that realize unambiguous state discrimination, one may
also consider the USD of sets of three or more linearly
independent states. Finally, let us emphasize that our approach of
choosing suitable cost functions and applying them to the dephased
quantum states might be as well useful for finding bounds on the
efficiency of implementing POVMs with linear optics.

\section*{Acknowledgments}

\noindent WM, PR, and NL acknowledge the support from the EU
project RAMBOQ. PvL and KN acknowledge funding from MIC in Japan.
PR and NL acknowledge the support of the DFG under the
Emmy-Noether programme. This work was also supported by the
network of competence QIP of the state of Bavaria (A8).

\appendix

\section{One-photon signal states}\label{onephoton}

Let us consider all those POVMs where the signal states contain
only {\it one photon}. In this typical and important case, any
unitary operation (gate) can be accomplished with linear optics
\cite{Reck}. This statement applies to arbitrary qu$d$it states,
where each basis vector of the qu$d$it is described by one photon
occupying one of $d$ modes, $\hat a_i^\dagger|{\bf 0}\rangle$,
$i=1...d$ (``multiple-rail encoding''). Similarly, any POVM can be
implemented solely by means of linear optics for these one-photon
signal states. This can be understood by looking at the
corresponding Naimark extension of the POVM. The POVM is then
described by a von Neumann measurement onto the orthogonal set
\begin{equation}\label{Naimark2}
|w_\mu\rangle = |u_\mu\rangle + |N_\mu\rangle \;,
\end{equation}
in a Hilbert space larger than the original signal space. Here,
the $\{|u_\mu\rangle\}$ are (unnormalized, potentially
nonorthogonal) state vectors in a Hilbert space $\cal K$ such that
\begin{equation}\label{POVMoperators2}
\hat E_\mu = |u_\mu\rangle\langle u_\mu |
\end{equation}
are the POVM operators of an $N$-valued POVM, $\mu=1...N$, with
$\sum_\mu\hat E_\mu=\mbox{1$\!\!${\large 1}}$. The vectors
$\{|N_\mu\rangle\}$ are defined in the complementary space $\cal
K^\bot$ orthogonal to $\cal K$, with the total Hilbert space $\cal
H = \cal K \oplus \cal K^\bot$. If the dimension of the signal
space is $n$, we have $|N_\mu\rangle = \sum_{i=n+1}^N b_{\mu
i}|v_i\rangle$ with some complex coefficients $b_{\mu i}$ and
$\{|v_i\rangle\}$ a basis in $\cal K^\bot$. In the multiple-rail
encoding, this leads to an orthogonal set of vectors
\begin{equation}\label{singleph1}
|w_\mu\rangle = \sum_{j=1}^N U_{\mu j} \hat a_j^\dagger|{\bf
0}\rangle\;,
\end{equation}
with a unitary $N\times N$ matrix $U$ having elements $U_{\mu j}$.
The application of a linear-optics transformation $V$ to this set
(in order to project onto it) can be written as
\begin{eqnarray}\label{singleph2}
|w_\mu\rangle \longrightarrow |w_\mu '\rangle&=& \sum_{j,k=1}^N
U_{\mu j} V^*_{k j}\hat a_k^\dagger|{\bf 0}\rangle
\nonumber\\
&=&\sum_{k=1}^N \delta_{\mu k} \hat a_k^\dagger|{\bf 0}\rangle =
\hat a_\mu^\dagger|{\bf 0}\rangle\;,
\end{eqnarray}
choosing $V\equiv U$. As a result, when detecting the outgoing
state, for every one-photon click in mode $\mu$, one can
unambiguously identify the input state $|w_\mu\rangle$.
This is why it is no surprise that any POVM for one-photon states
can be implemented via linear optics with unit success probability
(there is also an extension of this result for one-photon
implementations from any POVM to any Kraus operator
\cite{Ahnert,Ahnert2}).

For states other than one-photon states, it is a priori not clear
whether a given POVM can be implemented with linear optics. A
possible approach to deciding this would be to apply the criteria
for projective measurements \cite{PvLNorbert} to the orthogonal
set in Eq.~(\ref{Naimark}). The main difficulty then is that one
must consider any possible Naimark extension vectors
$\{|N_\mu\rangle\}$ in order to be able to decide whether the POVM
can be implemented or not. In particular, the extension of the
signal Hilbert space can be arbitrarily large. Therefore, in an
optical implementation, arbitrary ancilla states must be taken
into account, including arbitrarily many extra modes and photons.
It seems that, in general, more complicated approaches are
required to deal with the potentially infinite-dimensional problem
of adding arbitrary auxiliary states \cite{Jens04} (however, see
\cite{Knill1}). In this paper, we propose a dephasing approach to
the problem of implementing POVMs via linear optics, independent
of the Naimark extension.

\section{Alternative derivation of the optimal-USD fidelity
criterion}\label{rigorous}

Without referring to the general principle in
Eq.~(\ref{generalprinciple}) for arbitrary cost functions and
POVMs, here we directly derive the corresponding (necessary)
criterion for the special case of optimal USD in terms of
fidelities.

In general, for any state discrimination scheme based on {\it
static} linear optics, we have the following fidelity bounds,
\begin{eqnarray}\label{fidelitybounds}
F(\hat\rho_{+},\hat\rho_{-})\leq
F(\hat\rho_{+,H}',\hat\rho_{-,H}')\leq \left({\rm Prob}_{\rm
fail}^{\rm lin.opt.}\right)^2 .
\end{eqnarray}
In words, the fidelity of the linearly transformed and dephased
output states is lower bounded by the fidelity of the input states
and upper bounded by the squared failure probability in the
linear-optics implementation of unambiguous state discrimination.
The lower bound here corresponds to the general rule that the
fidelity of two density matrices cannot decrease under CPTP maps
\cite{Nielsenbook}. As for the upper bound, we may note that in
any scheme, the linearly transformed and dephased output states
take on the form of Eq.~(\ref{dephasedstatesafterlin})
corresponding to a total dephasing of the states in
Eq.~(\ref{nonorthogstatesafterlin}). Since the two density
matrices in Eq.~(\ref{dephasedstatesafterlin}) are diagonal in the
Fock basis and commute, we have the relation in
Eq.~(\ref{fidafterlinoptanddephase}). Now the failure probability
is given by ${\rm Prob}_{\rm fail}^{\rm lin.opt.} = \sum_m({\rm
P}^+_{m} + {\rm P}^-_{m})/2$. However, we also have $({\rm
P}^+_{m} + {\rm P}^-_{m})/2 \geq \sqrt{{\rm P}^+_{m}{\rm
P}^-_{m}}$, $\forall m$, thus proving the upper bound in
Eq.~(\ref{fidelitybounds}).

According to the fidelity bounds in Eq.~(\ref{fidelitybounds}), we
obtain Eq.~(\ref{generalrule}) as a {\it necessary condition} for
the optimal USD of two states via static linear optics and photon
counting, because optimal USD requires $\left({\rm Prob}_{\rm
fail}^{\rm lin.opt.}\right)^2 =
\left|\langle\chi_{+}|\chi_{-}\rangle\right|^2 =
F(\hat\rho_{+},\hat\rho_{-})$.

One can now further exploit the fact that the bounds in
Eq.~(\ref{fidelitybounds}) also hold for partially dephased
density matrices, corresponding to schemes that include {\it
conditional dynamics}. In particular, the upper bound in
Eq.~(\ref{fidelitybounds}) holds for the partially dephased
density matrices as well, because in any mixed-state
discrimination scheme, the squared failure probability is lower
bounded by the fidelity of the mixed states \cite{recentPhilippe}.

\section{Alternative derivation of the USD conditions for a fixed
array}\label{alternative}

Let us consider the optimal USD of two pure nonorthogonal states
using a {\it fixed array} of linear optics. We will give an
alternative derivation of the conditions in
Eq.~(\ref{nonorthogonalstatenewhierarchy}) and
Eq.~(\ref{nonorthogonalstatenewhierarchymodulus}), independent of
the fidelity criterion in Eq.~(\ref{generalrule}).

After the linear-optics transformation, the output states will
always take on the form of Eq.~(\ref{nonorthogstatesafterlin}),
for convenience, written again here,
\begin{eqnarray}\label{nonorthogstatesafterlinappendix}
|\chi_{+,H}\rangle &=& \sum_{k}\alpha_{k} |\{k\}\,\rangle
+\sum_{m}\alpha_{m} |\{m\}\,\rangle , \nonumber\\
|\chi_{-,H}\rangle &=& \sum_{l}\beta_{l} |\{l\}\,\rangle +
\sum_{m}\beta_{m} |\{m\}\,\rangle \,.
\end{eqnarray}
The patterns labeled by $k$ and $l$ are those that unambiguously
refer to the $+$ state and to the $-$ state, respectively. Because
of the finite overlap of the input states, we must include
patterns that occur in the expansion of both states. These
ambiguous patterns are denoted by the index $m$. In general, the
amplitudes of the ambiguous $N$-mode Fock states in the
expansions, and hence the probabilities for the corresponding
patterns to be detected, may be different for the $+$ and the $-$
state. In the following, we will first prove that in any {\it
optimal} USD scheme, the modulus of the amplitudes of any failure
pattern must indeed be equal for both states. Further, we will
show that for optimal USD, any relative phases in the expansion of
the failure patterns are reduced to a single global phase. As a
result, the output states after linear optics in optimal USD must
be describable in a three-dimensional vector space such that
\begin{eqnarray}\label{nonorthogstatesafterlin3dim}
|\chi_{+,H}\rangle &=& |\psi^+_{\rm concl}\rangle +
|\psi_{\rm inconcl}\rangle , \nonumber\\
|\chi_{-,H}\rangle &=& |\psi^-_{\rm concl}\rangle +
e^{i\phi}\,|\psi_{\rm inconcl}\rangle \,.
\end{eqnarray}
Here the states $|\psi^+_{\rm concl}\rangle$, $|\psi^-_{\rm
concl}\rangle$, and $|\psi_{\rm inconcl}\rangle$ are all mutually
orthogonal. They represent the vectors of all conclusive patterns
for the $+$ state, of those for the $-$ state, and the vector of
all inconclusive patterns, respectively.

As for the proof, we exploit the fact that in optimal USD, the
failure probability equals the modulus of the overlap of the
states to be discriminated, ${\rm Prob}_{\rm fail} = |\langle
\chi_{+}|\chi_{-}\rangle |$ (assuming equal a priori
probabilities). This implies that a linear-optical implementation
of optimal USD must satisfy
\begin{eqnarray}\label{proof1}
{\rm Prob}_{\rm fail}^{\rm lin.opt.} &=& \frac{1}{2}\, \sum_m
(|\alpha_m|^2 + |\beta_m|^2) \stackrel{!}{=}
|\langle \chi_{+}|\chi_{-}\rangle |\nonumber\\
&=& |\langle \chi_{+,H}|\chi_{-,H}\rangle | = \left| \sum_m
\alpha_m^* \beta_m \right| \,,
\end{eqnarray}
using Eq.~(\ref{nonorthogstatesafterlinappendix}). The factor
$1/2$ in the first line of Eq.~(\ref{proof1}) corresponds to the a
priori probabilities. Then, because of $\left| \sum_m \alpha_m^*
\beta_m \right|\leq\sum_m |\alpha_m^* \beta_m |$, we also have
\begin{eqnarray}\label{proof2}
{\rm Prob}_{\rm fail}^{\rm lin.opt.} &=& \frac{1}{2}\, \sum_m
(|\alpha_m|^2 + |\beta_m|^2)  \nonumber\\
&\leq& \sum_m |\alpha_m^* \beta_m | \,,
\end{eqnarray}
or,
\begin{eqnarray}\label{proof3}
\sum_m (|\alpha_m| - |\beta_m|)^2 \leq 0  \,.
\end{eqnarray}
The last inequality proves that $|\alpha_m| = |\beta_m|$, $\forall
m$. Moreover, it implies that $\left| \sum_m \alpha_m^* \beta_m
\right|\stackrel{!}{=}\sum_m |\alpha_m^* \beta_m |$, and hence
\begin{eqnarray}\label{proof4}
\left|\sum_m |\alpha_m|^2 e^{i\phi_m}\right| \stackrel{!}{=}
\sum_m |\alpha_m|^2 \,,
\end{eqnarray}
using $\beta_m = \alpha_m e^{i\phi_m}$. However,
Eq.~(\ref{proof4}) can only be satisfied for $e^{i\phi_m} =
e^{i\phi}$, $\forall m$. This concludes the proof of
Eq.~(\ref{nonorthogstatesafterlin3dim}). For the case of optimal
USD, we can now replace
Eq.~(\ref{nonorthogstatesafterlinappendix}) by
\begin{eqnarray}\label{nonorthogstatesafterlinoptimalUSD}
|\chi_{+,H}\rangle &=& \sum_{k}\alpha_{k} |\{k\}\,\rangle
+\sum_{m}\alpha_{m} |\{m\}\,\rangle , \nonumber\\
|\chi_{-,H}\rangle &=& \sum_{l}\beta_{l} |\{l\}\,\rangle +
e^{i\phi} \sum_{m}\alpha_{m} |\{m\}\,\rangle \,.
\end{eqnarray}
Let us now use this result in order to calculate the first-order
expression. Similar to
Eq.~(\ref{nonorthogonalstate1storderfromfid}), we obtain now
\begin{eqnarray}\label{nonorthogonalstate1storder}
\langle\chi_{+}|\hat c^\dagger_j \hat c_j|\chi_{-}\rangle &=&
\langle\chi_{+,H}|\hat a^\dagger_j \hat a_j|\chi_{-,H}\rangle
\\
&=& e^{i\phi} \sum_m |\alpha_m|^2  \langle \{m\}\,|\hat
a^\dagger_j \hat a_j|\{m\}\,\rangle \,.\nonumber
\end{eqnarray}
Since analogous expressions can be found for all higher orders,
the same arguments as those in the discussion after
Eq.~(\ref{nonorthogonalstate1storderfromfid}) apply here again.
Thus, finally we obtain the same hierarchies of {\it necessary}
conditions as in Eq.~(\ref{nonorthogonalstatenewhierarchy}) and
Eq.~(\ref{nonorthogonalstatenewhierarchymodulus}) for optimal USD
using a fixed array of linear optics.

For the special case of exact discrimination of two orthogonal
states $|\chi_{+}\rangle$ and $|\chi_{-}\rangle$ via photon
counting, the linearly transformed states take on the form,
\begin{eqnarray}\label{orthogonalstatesafterlin}
|\chi_{+,H}\rangle &=& \sum_{k}\alpha_{k} |\{k\}\,\rangle ,
\nonumber\\
|\chi_{-,H}\rangle &=& \sum_{l}\beta_{l} |\{l\}\,\rangle  \,.
\end{eqnarray}
Now there are no ambiguous patterns in the expansions. Let us
again examine the expression
\begin{eqnarray}\label{orthogonalstate1storder}
\langle\chi_{+}|\hat c^\dagger_j \hat c_j|\chi_{-}\rangle =
\langle\chi_{+,H}|\hat a^\dagger_j \hat a_j|\chi_{-,H}\rangle \,.
\end{eqnarray}
According to Eq.~(\ref{orthogonalstatesafterlin}), the output
states of the linear-optics transformation in exact state
discrimination must satisfy $\langle\chi_{+,H}|\hat a^\dagger_j
\hat a_j|\chi_{-,H}\rangle = 0$, because annihilating a photon in
the $j$th mode of the two states only yields a nonzero overlap for
coinciding patterns. Similarly, we have
\begin{eqnarray}\label{orthogonalstateanyorder}
\langle\chi_{+}|\hat c^\dagger_j\hat c^\dagger_{j'} \hat
c^\dagger_{j''}\cdots\hat c_j\hat c_{j'} \hat
c_{j''}\cdots|\chi_{-}\rangle &=&
\\
\langle\chi_{+,H}|\hat a^\dagger_j\hat a^\dagger_{j'} \hat
a^\dagger_{j''}\cdots\hat a_j\hat a_{j'} \hat
a_{j''}\cdots|\chi_{-,H}\rangle &=& 0\,,\,\forall j,j',j''...\,,
\nonumber
\end{eqnarray}
because annihilating a photon in the $j$th, $j'$th, $j''$th, etc.,
mode of the two states also only yields a nonzero overlap for
coinciding patterns. Thus, we end up having the following set of
conditions for exact state discrimination
\begin{eqnarray}\label{orthogonalhierarchy1}
\langle\chi_{+,H}|\hat a^\dagger_j\hat a_j |\chi_{-,H}\rangle &=&
0\,,
\quad\forall j\;,\\
\langle\chi_{+,H}|\hat a^\dagger_j\hat a^\dagger_{j'} \hat a_j\hat
a_{j'} |\chi_{-,H}\rangle &=& 0\,,
\quad\forall j,j'\;, \nonumber\\
\langle\chi_{+,H}|\hat a^\dagger_j\hat a^\dagger_{j'} \hat
a^\dagger_{j''}\hat a_j\hat a_{j'} \hat a_{j''} |\chi_{-,H}\rangle
&=& 0\,,
\quad\forall j,j',j''\;, \nonumber \\
\quad\quad\quad\vdots\quad\quad\quad &=&\quad\quad\quad \vdots
\nonumber
\end{eqnarray}
or, equivalently, as described in
Eq.~(\ref{orthogonalhierarchy2}).

\section{Derivation of the USD conditions for conditional dynamics}
\label{conditionaldynamics}

We consider the optimal USD of two pure nonorthogonal states via
linear optics including {\it conditional dynamics}. Let us assume,
without loss of generality, that mode 1 is detected first,
corresponding to a partial dephasing of the states only with
respect to that mode. Now instead of writing the states after
linear optics as in Eq.~(\ref{nonorthogstatesafterlin}), we use
the following expressions,
\begin{eqnarray}\label{nonorthogstatesafterlinconddyn}
|\chi_{+,H}\rangle &=& \sum_{k}\alpha_{k} |k\rangle_1\otimes
|\tilde\gamma_k^+\rangle +\sum_{m}\alpha_{m} |m\rangle_1 \otimes
|\tilde\gamma_m^+\rangle , \nonumber\\
|\chi_{-,H}\rangle &=& \sum_{l}\beta_{l} |l\rangle_1\otimes
|\tilde\gamma_l^-\rangle +\sum_{m}\beta_{m} |m\rangle_1 \otimes
|\tilde\gamma_m^-\rangle \,,\nonumber\\
\end{eqnarray}
where this time, the states $|k\rangle_1$ and $|l\rangle_1$
represent those number states of mode 1 which only occur in the
expansion of the $+$ and the $-$ state, respectively. The one-mode
states $|m\rangle_1$ lead to the ambiguous detection events in
mode 1. Finally, the states $|\tilde\gamma_k^+\rangle$, etc.,
refer to the corresponding conditional states of the remaining
modes (after normalization). Similarly, for the partially dephased
density operators, we obtain
\begin{eqnarray}\label{nonorthogstatesafterlinconddyndephase}
\hat\rho_{+,H}' &=& \sum_{k}{\rm P}^+_{k} |k\rangle_1\langle
k|\otimes|\tilde\gamma_k^+\rangle\langle\tilde\gamma_k^+|\nonumber\\
&+&\sum_{m}{\rm P}^+_{m} |m\rangle_1\langle m|
\otimes|\tilde\gamma_m^+\rangle\langle\tilde\gamma_m^+| , \nonumber\\
\hat\rho_{-,H}' &=& \sum_{l}{\rm P}^-_{l} |l\rangle_1\langle
l|\otimes|\tilde\gamma_l^-\rangle\langle\tilde\gamma_l^-|\nonumber\\
&+&\sum_{m}{\rm P}^-_{m} |m\rangle_1\langle
m|\otimes|\tilde\gamma_m^-\rangle\langle\tilde\gamma_m^-| \,.
\end{eqnarray}
Note that the partially dephased states are no longer diagonal in
the Fock basis, i.e., the conditional density matrices may contain
off-diagonal terms. The corresponding fidelities are now
\begin{eqnarray}
F(\hat\rho_{+},\hat\rho_{-}) &=& F(\hat\rho_{+,H},\hat\rho_{-,H})
\\
&=& \sum_{m,n} \alpha_m^*\alpha_n\beta_m\beta_n^*\,
\langle\tilde\gamma_m^+|\tilde\gamma_m^-\rangle
\langle\tilde\gamma_n^+|\tilde\gamma_n^-\rangle^*\,, \nonumber
\end{eqnarray}
and
\begin{eqnarray}
F(\hat\rho_{+,H}',\hat\rho_{-,H}') = \left(\sum_m\sqrt{{\rm
P}^+_{m}{\rm
P}^-_{m}}\left|\langle\tilde\gamma_m^+|\tilde\gamma_m^-\rangle
\right|\right)^2.
\end{eqnarray}
Finally, we end up having the following condition due to the
fidelity criterion in Eq.~(\ref{generalrule}),
\begin{eqnarray}
\sum_{m,n}\sqrt{{\rm P}^+_{m}{\rm P}^+_{n}{\rm P}^-_{m}{\rm
P}^-_{n}}\left|\langle\tilde\gamma_m^+|\tilde\gamma_m^-\rangle
\right|\left|\langle\tilde\gamma_n^+|\tilde\gamma_n^-\rangle
\right| \stackrel{!}{=} \nonumber\\
\sum_{m,n} \alpha_m^*\alpha_n\beta_m\beta_n^*
\langle\tilde\gamma_m^+|\tilde\gamma_m^-\rangle
\langle\tilde\gamma_n^+|\tilde\gamma_n^-\rangle^*\,,
\end{eqnarray}
or, in terms of the unnormalized conditional states,
\begin{eqnarray}
\sum_{m,n}\left|\langle\gamma_m^+|\gamma_m^-\rangle
\right|\left|\langle\gamma_n^+|\gamma_n^-\rangle \right|
\stackrel{!}{=} \sum_{m,n} \langle\gamma_m^+|\gamma_m^-\rangle
\langle\gamma_n^+|\gamma_n^-\rangle^*.
\end{eqnarray}
Now the only way to satisfy this condition is through
\begin{eqnarray}\label{overlapcondition}
\frac{\langle\gamma_m^+|\gamma_m^-\rangle}{\left
|\langle\gamma_m^+|\gamma_m^-\rangle\right|} \stackrel{!}{=}
\frac{\langle\gamma_n^+|\gamma_n^-\rangle}{\left
|\langle\gamma_n^+|\gamma_n^-\rangle\right|}\,,
\end{eqnarray}
for any nonzero overlaps labeled by $m$ and $n$. In other words,
for any two inconclusive one-mode detection events, any
non-vanishing overlaps of the conditional states coming from the
$+$ signal and the $-$ signal must have equal phases. Finally, we
can now again examine the first-order condition of our criteria,
however, here only for the detected mode 1,
\begin{eqnarray}\label{nonorthogonalstate1storderfromfidconddyn}
\langle\chi_{+}|\hat c^\dagger_1 \hat c_1|\chi_{-}\rangle &=&
\langle\chi_{+,H}|\hat a^\dagger_1 \hat a_1|\chi_{-,H}\rangle
\\
&=& \sum_m \langle\gamma_m^+|\gamma_m^-\rangle {\,\,_1}\langle
m|\hat a^\dagger_1 \hat a_1|m\rangle_1 \,.\nonumber
\end{eqnarray}
Similar expressions hold for the higher orders in mode 1. Note
that the different orders here are evaluated only for the first
mode to be detected, corresponding to the first step in a
conditional-dynamics scheme. Of course, in addition, one could
calculate further expressions using the conditional states of
modes 2 through $N$ in order to derive more criteria for a
conditional-dynamics protocol. Here, we only focus on the first
step in any conditional-dynamics scheme, namely the detection of a
first mode.

Using Eq.~(\ref{nonorthogonalstate1storderfromfidconddyn}) and
Eq.~(\ref{overlapcondition}) (for any nonzero overlaps), it
becomes clear now that the hierarchies of conditions necessary for
optimal USD when detecting a first mode $j$ are simply the subset
of the fixed-array conditions in
Eq.~(\ref{nonorthogonalstatenewhierarchy}) and
Eq.~(\ref{nonorthogonalstatenewhierarchymodulus}) referring to
this one mode. This subset of conditions is given in
Eq.~(\ref{nonorthogonalstatenewhierarchyconddyn}) and
Eq.~(\ref{nonorthogonalstatenewhierarchymodulusconddyn}).


\begin{thebibliography}{99}
\bibitem{footnote1} The statement in
Eq.~(\ref{generalprinciple}) can also be expressed in terms of
``score functions" $S$, where $S = 1-C$. An example for a possible
score function is the mean fidelity in quantum state estimation.
For the quantum mechanically optimal estimation of a given quantum
state drawn from a continuous or discrete set of states, the
``score" is then expressed in terms of the mean value of the
fidelity $F(\hat\rho_\mu,\hat\rho)$ between the actual input state
$\hat\rho$ and the estimated state $\hat\rho_\mu$ depending on the
result of a POVM $\hat E_\mu$. This mean value of
$F(\hat\rho_\mu,\hat\rho)$ can be calculated via $\bar F =
\sum_{\mu=1}^N\left\langle {\rm Tr}(\hat
E_\mu\hat\rho)F(\hat\rho_\mu,\hat\rho)\right\rangle$, where ${\rm
Tr}(\hat E_\mu\hat\rho)$ is the probability for obtaining the
result $\mu$ when the actual input state was $\hat\rho$, and
$\langle\cdots\rangle$ denotes the average over the entire
alphabet of input states. An optimal estimation scheme achieves
the maximum average fidelity $\bar F$, maximized with respect to
the chosen strategy given by the set $\{\hat
E_\mu,\hat\rho_\mu\}_{\mu=1}^N$.
\bibitem{footnote2} Another optimal POVM solution to estimating
an unknown qubit state is the covariant POVM that consists of a
continuous set of projectors on any possible qubit state
(projecting on the ``spin-coherent states"). Since this POVM
solution is infinite, it is not very practical for an
implementation. However, note that there are infinite POVMs that
can be implemented. An example is the optimal estimation of an
unknown coherent state $|\alpha\rangle$. The elements of the
optimal infinite POVM solution in this case are the coherent-state
projectors $|\beta\rangle\langle\beta|$. Its implementation only
requires a symmetric beam splitter and two homodyne detectors,
corresponding to a heterodyne detection or a measurement of the Q
function of the input state. The Naimark extension here is due to
the extra vacuum mode at the second input port of the beam
splitter.
\bibitem{footnote3} Even for more than two input states,
quantum mechanically, such a POVM
always exists, i.e., there is a nonzero probability for a
conclusive measurement result, provided the nonorthogonal input
states are linearly independent \cite{TonyPhysLett}. Conversely, a
given POVM is equivalent to the discrimination of linearly
independent signal states, if the POVM elements can be written as
rank one projectors onto the reciprocal vectors of the signal
states \cite{Tonyprivate}.
\bibitem{footnote4} Obviously, for only partially dephased density
matrices describing schemes that potentially include conditional
dynamics where only a first mode is detected, the fidelity
criterion in Eq.~(\ref{generalrule}) cannot be sufficient for
optimal USD.
\end{thebibliography}
\end{document}